\documentclass[onecolumn]{elsart}   

\usepackage{epsfig}
\usepackage{amsmath}                      

\def\lsim{\lower.5ex\hbox{$\; \buildrel < \over \sim \;$}}
\def\gsim{\lower.5ex\hbox{$\; \buildrel > \over \sim \;$}}
\def\g{\ifmmode \gamma \else $\gamma$\fi}
\def\ee{\end{equation}}
\def \be{\begin{equation}}
\def \ul{\underline}

\newcommand{\mathd}{\mathrm{d}}
\newcommand{\tmop}[1]{\operatorname{#1}}
\newcommand{\tmmathbf}[1]{\boldsymbol{#1}}
\newcommand{\mathe}{\mathrm{e}}
\renewcommand{\vec}[1]{\boldsymbol{#1}}
\begin{document}

\begin{frontmatter}

\title{Gravitational Radiation from \\ Ultra High Energy Cosmic Rays\\
in Models with Large Extra Dimensions} 

\author[Koch]{Ben Koch}\ead{koch@th.physik.uni-frankfurt.de},    % Add the 
\author[Koch,Bleicher]{Hans-Joachim Drescher}
\author[Bleicher]{, Marcus Bleicher}

\address[Koch]{Frankfurt Institute for Advanced Studies (FIAS),\\
60438 Frankfurt am Main, Germany}                            
\address[Bleicher]{Institut f{\"u}r Theoretische Physik, 
Johann Wolfgang Goethe-Universit{\"a}t,\\
60438 Frankfurt am Main, Germany}             % full addresses


\begin{abstract}                        
The effects of classical gravitational radiation in models with large extra dimensions
are investigated for ultra high energy cosmic rays (CRs).
The cross sections are implemented into a simulation package (SENECA) for high energy
hadron induced CR air showers.  
We predict that gravitational radiation from quasi-elastic scattering
could  be observed at incident CR energies above $10^9$~GeV for a setting with 
more than two extra dimensions.
It is further shown that this gravitational energy loss can 
alter the energy reconstruction for CR energies $E_{\rm CR}\ge 5\cdot 10^9$~GeV.
\end{abstract}

\end{frontmatter}

%%%%%%%%%%%%%%%%%%%%%%%%%%%%%%%%%%%%%%%%%%%%%%%%%%%%%%%%%%%%%

\section{Motivation}

One of the major problems in modern physics is to
combine quantum physics and gravitation. The most promising candidate to
solve this problem seems to be string theory because it includes all
symmetry groups of the standard model, however, it introduces additional 
space dimensions \cite{Green}. From the present (non-)observation of
these dimensions it is concluded that they are compactified and their size was assumed 
to be on the Planck scale. However, string theory itself does not give any
stringent constraints on the size of the extra dimensions other than
non-observability so far. 

This opens a possibility for gravity-only extra dimensions,
assuming that only gravitons are able to enter these additional
dimensions. The size of these dimensions is then primarily constraint
by direct measurements of the gravitational inverse square law.
However, the strength of gravitational interactions has only been
measured down to a scale of some micrometers
\cite{Hoyle:2004cw,Adelberger:2003zx}. Below this scale some
modification of the gravitational force law is still possible.
This allows to introduce a new fundamental scale of gravity $M_f\ll M_{Planck}$ in the TeV-range. 
These Large Extra Dimensions models (LXDs) have compactification radii up to $\sim \mu$m 
and a new fundamental scale in the TeV range \cite{Arkani-Hamed:1998rs,Arkani-Hamed:1998nn,Antoniadis:1998ig}. 
If the new scale of gravity is indeed not far beyond the electro-weak scale, it might be reachable
in future collider experiments at LHC and beyond or in cosmic ray observatories like AUGER or ICECUBE. 
At center of mass energies above the new fundamental scale $M_f$ the LXDs become important
and first observable effects of quantum gravity might be observable.
A multitude of new effects have been predicted in the recent years to obtain deeper insight into
this exciting field ranging from Kaluza-Klein graviton production and increased neutrino cross sections to black hole production \cite{Argyres:1998qn,Giudice:1998ck,Han:1998sg,Mirabelli:1998rt,Vacavant:sd,Banks:1999gd,Atwood:1999cy,Atwood:1999pn,Atwood:1999zg,Rizzo:1999pc,Ahern:2000jn,Cullen:2000ef,Giddings:2001ih,Hossenfelder:2001dn,Hofmann:2001pz,Bleicher:2001kh,Rizzo:2001ag,Dvergsnes:2002nc,Hossenfelder:2003dy,Hossenfelder:2003jz,Buanes:2004ya}. 

Apart from these promising phenomena, present day observations also allow
for direct and indirect constraints on the new fundamental scale and the size of
the LXDs:
\begin{itemize}
\item
A fundamental
scale in the TeV-range does not allow a single LXD as the compactification
radius would get in the order of the size of our solar system and we would
have noticed. 

\item
For higher numbers of LXDs the compactification radius does
not exceed the $\mu$m-scale as shown by the mentioned short distance
experiments.

\item
Even smaller distances can be probed in collider experiments. At
the moment the highest center of mass energies are reached in hadron-hadron
collisions at Tevatron with 2 TeV cm-energy which sets a bound of $M_f\ge
1.0$ TeV for 2 down to 0.7 TeV for 6 LXDs \cite{Abbott:2000zb,Acosta:2003tz}.

\item
Other bounds are set by supernova explosions
\cite{Cullen:1999hc,Hannestad:2003yd,Hanhart:2000er,Hanhart:2001fx}. The
cooling of the supernovae is modified by the production of gravitons
which is very sensitive to the number of LXDs in the reached energy
region. For 2 LXDs a bound of about 500 TeV is obtained. However, for
any higher number of LXDs the bounds are lower than from the collider
experiments.
\end{itemize}

In the last years it was recognized that cosmic rays (CRs) provide an excellent 
laboratory to study the onset of physics beyond
the standard model. The initial  energies of UHECRs exceeds 
$10^{11}$~GeV ($\sqrt{s_{\rm NN}}\sim 100-1000$~TeV) and might therefore
allow one to  probe TeV scale large extra
dimensions \cite{Domokos:1998ry,Emparan:2001kf,Ringwald:2001vk,Anchordoqui:2001cg,Kazanas:2001ep,Kowalski:2002gb,Alvarez-Muniz:2002ga,Sigl:2002bb,Anchordoqui:2003jr,Anchordoqui:2004ma,Anchordoqui:2005gj,Anchordoqui:2005is}. 
Especially the radiation of  gravitational waves in models with large extra dimensions is strongly enhanced
compared to standard general relativity. As will be discussed later, this might lead to 
observable signatures and modifications of the reconstructed flux and energy estimates for UHECRs. 

To explore the effect of gravitational energy loss in cosmic ray air-showers, we use the differential cross section
for gravitationally radiated energy in quasi-elastic scatterings as calculated in
\cite{Cardoso:2002pa,Koch:2005bc} for $2 \rightarrow 2$ processes.  Then we
apply the result to cosmic ray air shower simulations and extract the
impact on cosmic ray observables.

%%%%%%%%%%%%%%%%%%%%%%%%%%%%%%%%%%%%%%%%%%%%%%%
\section{Gravitational radiation from quasi-elastic scattering with extra dimensions}
\label{sec_XDrad}

First estimates to study effects of gravitational energy loss of CRs due to the presence of 
extra dimensions were explored by \cite{Kazanas:2001ep}. There, the presence of large 
extra dimensions was incorporated into the well known results from general relativity \cite{Weinberg:1972} by a change 
of the phase space seen by the emitted gravitational wave.  
The additional phase space factor for the emitted gravitational wave was given by 
\begin{equation}\label{phasespace}
g_{d}(k_d)=
\frac{(k_d R)^{d}}
{d \Gamma(d/2) \pi^{d/2} 2^{d-1}}.
\end{equation}
Note that $g_0=1$.
Where $R$ is the compactification radius of the extra dimensions
in the ADD scenario given by \cite{Arkani-Hamed:1998rs}:
\begin{equation}
R=M_f^{-\frac{d+2}{d}} M_{\rm Pl}^{\frac{2}{d}}.
\end{equation}
Here, $M_f$ is the new fundamental scale and $M_{\rm Pl}$ is the four dimensional Planck mass
related to the gravitational constant by $G_{\rm N}=1/M_{\rm Pl}^2$.

This method led to a strong modification of the reconstructed energy spectrum and the authors concluded that
the steepening of the CR energy spectrum around $10^{15.5}$~eV (the ''knee'') might be due to gravitational energy loss.

However, from our present  studies (see also  \cite{Cardoso:2002pa}) it seems that a calculation of the effects of the 
gravitational energy loss requires a more elaborate treatment as will be discussed now.
The simplified treatment can be improved by direct calculation\footnote{We will use the following notations:
$4+d$ space-time vectors will be $x=(x_0,\underline{x})$, where the spatial part
can be split again into a three dimensional
and a $d$ dimensional part $\underline{x}=(\vec{x},x_{\perp})$.}%
of  the gravitational energy loss 
in a $N \rightarrow M$ scattering process as given by \cite{Cardoso:2002pa,Koch:2005bc}:
\be\label{eq_dEdOmdom5}
%\begin{array}{rl}
\frac{dE}{d\Omega_{3+d} dk_0}
=
\frac{1}{M_f^{2+d}}\frac{k_0^{2+d}}{2(\pi)^2(2\pi)^d}
\sum\limits_{I,J}\frac{\eta_I\eta_J}{(P_{(I)}k)(P_{(J)}k)}
\left[
(P_{(I)}^{\mu}P_{(J)\mu})^2-\frac{1}{2+d}P_{(I)}^2\, P_{(J)}^2
\right],
%\end{array}
\ee
The $P_{(N)}$ are the momenta of the colliding particles
and the factors $\eta_N$ are defined by
\begin{equation}
\eta_N=
\left\{ 
\begin{array}{rl}
-1 & \mbox{\, for initial state particles.}\\
+1 & \mbox{\, for final state particles.}
\end{array}
\right.
\end{equation}
Thus, in the case of a $2 \rightarrow 2$ collision the index $N$ runs from 1 to 4.
Before we continue, we want to point out that  equation
(\ref{eq_dEdOmdom5}) follows from classical considerations and is not derived 
 from any form of quantum theory of gravity (e.g. loop quantum gravity, SUGRA or string theory).  
However, we believe that it can account - at least semi-quantitatively - for the major effects of
the  gravitational energy loss.

Next, we integrate Eq. (\ref{eq_dEdOmdom5}) with the help of the Mathematica package FeynCalc
\cite{Mertig:1990an,Wolfram:1996}.  Difficulties for the
$d\Omega_{3+d}$ integration arise from the $P \cdot k$ terms in the
numerator. The protons are bound to the brane and the product $P \cdot k$ gives
for example for one of the incoming protons:
\begin{equation}
P_1 \cdot k = P^0_1 k_0- \vec{p_1} \vec{k}-0=
|\vec{p}|k_0\left(\sqrt{1+\frac{m_p^2}{|\vec{p}^2|}}-
\sqrt{1-\frac{k_d^2}{k_0^2}}\cos{\phi_k}\right).
\end{equation}
For ${k_d^2}/{k_0^2}\approx 0$ and $\phi_k\approx 0$,
$P_1 \cdot k$ becomes small and the denominator in Eq. (\ref{eq_dEdOmdom5})
is only regularized by  ${m_p^2}/{|\vec{p}^2|}$.

We introduce the Mandelstam variables $s$, $t$ and $u$ by
\be
s=(P_1+P_2)^2,\quad t=(P_1-P_3)^2,\quad u=(P_1-P_4)^2.
\ee
It is convenient to perform  a coordinate transformation to rewrite Eq. (\ref{eq_dEdOmdom5})
in terms of spherical coordinates in three dimensional space and 
the $d$ extra-dimensional coordinates separately:
\begin{equation}
\frac{dE}{dk_0 d\Omega_{3+d}}=
\frac{k^{d+2}dE}{dk^{3+d}}=
\frac{k^{d+2}dE}{dk^{d}dk^{3}}=
\frac{(\vec{k}^2+k_d^2)^{(d+2)/2}dE}
{\vec{k}^2d\vec{k}d\Omega_3 |k_d|^{d-1}dk_d d\Omega_d}.
\end{equation}
Solving this for the new integration variables yields
\begin{equation}
\label{eqtrafo}
\frac{dE}{d\vec{k}d\Omega_3 dk_d d\Omega_d}=
\frac{dE}{dk_0 d\Omega_{3+d}}
\frac{\vec{k}^2k_d^{d-1}}{(\vec{k}^2+k_d^2)^{(d+2)/2}}.
\end{equation}
The first term on the right side can be approximated by
Eq. (\ref{eq_dEdOmdom5}) as soon as the wavelength
of the gravitational wave is smaller than the compactification
radius $R$ of the extra dimensions and the gravitational wave can propagate freely into
the bulk. Rephrased as a condition for $|k|$ this constrained becomes
\begin{equation}
\label{eqkcond}
|k|>M_f\left(
\frac{M_f^{2}}{M_{\rm Pl}^2}
\right)^{\frac{1}{d}}.
\end{equation}
A lower bound on $|k|$ is not relevant for the energy loss discussion, because the major contribution to
the radiated energy  comes from the high energy (i.e. large $|k|$) part.
To calculate the energy loss due to the gravitational wave emission one has to perform
the $d\Omega_3=\sin{\phi_{kz}}d\phi_k d\phi_{kz}$, the 
$d\Omega_d$, the $d\vec{k}$ and the $dk_d$ integrals.
However, the rather steep $t$ dependence of the elastic Proton-Proton cross section
allows us to simplify these integrals, because the physically relevant processes are dominated by 
small $|t| <m_p^2$ contributions, with $m_p$ being the mass of the Proton.
Thus, one can expand Eq. (\ref{eq_dEdOmdom5}) for small $|t|$. This gives
for the part $\sum_{I,J}\dots$ in Eq. (\ref{eq_dEdOmdom5}) containing 
the sums over external momenta
\begin{equation}
\label{SumXD}
\begin{array}{lll}
\sum\limits_{I,J}\dots &=& 
-8 t 
\left\{  {( k_0^2 -
          k_d^2         ) }^3 
     ( 4 
        m_p^2 - s
       )  s^4 
     {\cos ({{\phi_k}
         })}^6  \right.\\
 &+&
    k_0^2 
     ( k_0^2 - k_d^2
       )  s 
     ( 4 
        m_p^2 + s
       )  
     {\cos ({{\phi_k}
         })}^2 
     \left[ k_d^2 s 
        ( -8 
           m_p^4
           - 4 
           m_p^
          2 s + s^2 )\right.\\ 
  &-& 
	4 
        k_0^2 
        ( 32 
           m_p^6
           - 14 
           m_p^
          4 s - 
          3 
           m_p^
          2 s^2 + s^3 )
           \\ 
 &+&\left.
       ( k_0
          ^2 - 
          k_d^2
          )  s 
        ( -8 
           m_p^4
           - 4 
           m_p^
          2 s + s^2 ) \cos (2 
          {{\phi_k}})
       \right]\\
  &-&
    0.5 \left[{( 
           k_0^2 - 
           k_d^2
           ) }^2 s^2 
       {\cos ({{\phi_k}
          })}^4 
       \left( k_d
          ^2 s 
          ( 8 
           m_p^4
           - 4 
           m_p^
          2 s + s^2 ) \right.\right.\\
  &+& 
	k_0^
          2 
          ( -128 
           m_p^6
           + 88 
           m_p^
          4 s +
           12 
           m_p^
          2 s^2 - 7 s^3
           )\\
 & +& \left.\left.
         ( k_0
           ^2 - 
           k_d^2
           )  s 
          ( 8 
           m_p^4
           - 4 
           m_p^
          2 s + s^2 ) \,\cos (2 
           {{\phi_k}})
         \right)\right] \\
 &-&
    0.5\left[k_0^4 
       {( 4 
           m_p^2
           + s ) }^2 
       \left( k_d
          ^2 s 
          ( 8 
           m_p^4
           - 4 
           m_p^
          2 s + s^2 )\right.\right.\\ 
 &+& 
k_0^
          2 
          ( -128 
           m_p^6
           + 24 
           m_p^
          4 s  +
           12 
           m_p^
          2 s^2 - 3 s^3
           )  \\
&+& \left.\left.\left.
         ( k_0
           ^2 - 
           k_d^2
           )  s 
          ( 8 
           m_p^4
           - 4 
           m_p^
          2 s + s^2 )
           \cos (2 
           {{\phi_k}})
         \right) \right]\right\}\\
 &/&
	 \left\{k_0^8 
  \left( -4 
     {{m_p}}^2 + s
    \right)  
  {\left[ 4 
       m_p^2 + 
      s + \left( -1 + 
         \frac{k_d
          ^2}{k_0
           ^2} \right)  s 
       {\,\cos ({{\phi_k}
          })}^2 \right] }^4\right\}.
\end{array}
\end{equation}
For $k_d \approx 0$
the radiation does not propagate into the extra dimensions
and Eq. (\ref{SumXD}) reduces to the well known classical limit.
From Eq. (\ref{SumXD}) one can see that for
$k_d^2/k_0^2 s\ge 4 m_p^2$ the regularising part
in the denominator is not $m_p^2/s$ any more and
a Taylor expansion of Eq. (\ref{SumXD}) around ${m_p^2}/{s}=0$ 
is allowed. This expansion has a large validity region
for ultra high energy collisions because it just demands that 
\begin{equation}
\label{eqkdgr}
\frac{\sqrt{s}}{2} > k_d^2 > \vec{k}^2 \frac{4 m_p^2}{s}.
\end{equation}
This approximation also fulfils the condition in Eq. (\ref{eqkcond}).
After performing the integration $d\phi_{kz}$, this series gives
\begin{equation}
\frac{dE}{dk_d d\vec{k} d\phi_k d\Omega_d}=
\frac{t}{(2\pi)^{d+1}M_f^{d+2}}
\frac{2 k_d^{d-1} \vec{k}^2\left[(k_0^2-k_d^2)\cos{(2\phi_k)}-k_0^2\right]}
{\left[k_0^2+(k_d^2-k_0^2)\cos{(\phi_k)}\right]^2}~.
\end{equation}
Next we perform the $\phi_k$ integration,
\begin{equation}
\frac{dE}{dk_d d\vec{k} d\Omega_d}=
\frac{t}{(2\pi)^{d}M_f^{d+2}}
\frac{k_d^{d-2} \vec{k}^2 (2k_d^2+3\vec{k}^2)}
{(k_d^2+\vec{k}^2)^2} ~.
\end{equation}
The integration over
the $d$ dimensional unit sphere $\Omega_d$ gives a factor 
${2\pi^{d/2}}/{\Gamma(d/2)}$.
\begin{equation}
\frac{dE}{dk_d d\vec{k}}=
\frac{t}{2^{d-1}\pi^{d/2}\Gamma (d/2)M_f^{d+2}}
\frac{ k_d^{d-2} \vec{k}^2 (2k_d^2+3\vec{k}^2)}
{(k_d^2+\vec{k}^2)^2} ~.
\end{equation}
Next, the $k_d$ and the $|\vec{k}|$ integration can be performed with
respect to the integration limits $k_d^2+\vec{k}^2< k_{max}$ and
Eq. (\ref{eqkdgr}). This  calculation can be done explicitly for two, four and six
extra dimensions:
\be\label{eq_EvonTd246}
\begin{array}{rcl}
E(t,d=2)&=&-k_{max}^3 t \left[5\sqrt{2}-\log{(1+\sqrt{2})}\right]/(12 \pi M_f^4),\\
E(t,d=4)&=&
-k_{max}^5 t \left[\sqrt{2}-\log{(1+\sqrt{2})}\right]/(16 \pi^2 M_f^6),\\
E(t,d=6)&=&
-k_{max}^7 t \left[11\sqrt{2}-13\log{(1+\sqrt{2})}\right]/(1792 \pi^3
M_f^8).
\end{array}
\ee
Let us now  discuss the relation between this result and those obtained 
in earlier publications:
\begin{itemize}
\item
In Ref. \cite{Cardoso:2002pa} 
the gravitational wave was assumed to have a momentum vector only in the direction 
out of the brane, thus the denominator in Eq. (\ref{eq_dEdOmdom5})
simplifies to $P_I k = E^0_I k_0$. After integrating
over $k_0$ (which is not strictly correct, because the problem
is not spherically symmetric in $3+d$ spatial dimensions any more)
the result shows the same $t$ and $k_{max}$ dependence
as Eq. (\ref{eq_EvonTd246}). The different factors are due to the
simplification in the integration.
\item
The phase space argument used in Ref. \cite{Kazanas:2001ep} leads
to the same $k_{max}$ dependence. However, the pre-factors differ and
even more striking the result derived in \cite{Kazanas:2001ep}
has no $t$  dependence  (but $s$ instead). Therefore, this approach  leads to drastic
overestimation of the gravitational energy loss in high energy cosmic
rays, as we will see in the following sections.
\end{itemize}

%%%%%%%%%%%%%%%%%%%%%%%%%%%%%%%%%%%%%%%%%%%%%%%%%%%%%%%%

\section{Quasi-Elastic hadron-nucleus scattering}

In order to calculate the energy loss due to gravitational wave emission 
in air showers at high energies
one has to know the elastic scattering cross section $\mathd \sigma_{\rm elastic} / \mathd
t$. We construct it from the hadron-nucleon scattering cross section.

In the impact parameter representation, the profile function is defined as

\begin{equation}
    \Gamma ( s, b )  =  \frac{\sigma_{\tmop{\rm total}}}{2}  \frac{1}{2 \pi
    B_{\tmop{\rm elastic}}} \exp \left( - \frac{b^2}{2 B_{\tmop{\rm elastic}}} \right)\quad,
    \label{for:profile}
\end{equation}
where $B$ is the elastic scattering slope and $\sigma_{\rm total}$ is the total hadron-nucleon cross section. 
The profile function is related to the amplitude,
\begin{equation}
    \Gamma ( s, b )  =  - \frac{i}{8 \pi} A ( s, b )\quad.
\end{equation}
The profile function Eq. (\ref{for:profile}) can be expressed via the eikonal
\cite{Levy:1969cr} function $\chi$ by
\be
  \Gamma ( s, \tmmathbf{b} )  =  1 - \exp\left[i \chi ( \tmmathbf{b})\right]\quad, 
\ee

and  is related to the phase shift of the scattered wave.

The Fourier transform relates impact parameter space to momentum space by
\begin{equation}
    A ( s, t )  =  \frac{s}{4 \pi} \int \mathd^2  \tmmathbf{b}\, A ( s,
    \tmmathbf{b} ) \exp (- i\tmmathbf{b} \tmmathbf{q})\quad, \label{for:Ast}
\end{equation}
with $t=-q^2\approx-\tmmathbf{q}^2$.
The amplitude gives the differential elastic cross section:
\be
 \frac{\mathd \sigma_{\tmop{\rm elastic}}}{\mathd t}  = 
 \frac{1}{16 \pi s^2} |A ( s, t ) |^2 
  = \left. \frac{\mathd
  \sigma_{\tmop{\rm elastic}}}{\mathd t} \right|_{t = 0} \exp (
 B_{\tmop{\rm elastic}} t )\quad. \label{for:dsigdt}
\ee

From the last expression we can define the elastic scattering slope
\be
  B  \equiv  \left[ \frac{\mathd}{\mathd t} \left( \ln \frac{\mathd
  \sigma_{\tmop{\rm elastic}}}{\mathd t} \right) \right]_{t = 0} . \label{for:defB} 
\ee
Inserting Eqs. (\ref{for:Ast}) and (\ref{for:dsigdt}) into Eq. (\ref{for:defB}), the
scattering slope becomes
\be\label{eq_slope}
\begin{array}{rcl}
  B & = & \left. \frac{2 \frac{d}{\mathd t} | \int \mathd^2 \tmmathbf{b}A (
s,
  \tmmathbf{b} ) \mathe^{- i\tmmathbf{b} \tmmathbf{q}} |}{| \int \mathd^2
  \tmmathbf{b}A ( s, \tmmathbf{b} ) \mathe^{- i\tmmathbf{b} \tmmathbf{q}} |}
  \right|_{t = 0} \\
  & = & \left. \frac{2 \frac{\mathd}{- 2 q \mathd q} | \int \mathd^2
  \tmmathbf{b}A ( s, \tmmathbf{b} ) ( 1 + ( - i\tmmathbf{b} \tmmathbf{q}) +
  \frac{1}{2} ( - i\tmmathbf{b} \tmmathbf{q})^2 - \ldots ) |}{| \int
\mathd^2
  \tmmathbf{b}A ( s, \tmmathbf{b} ) |} \right|_{t = 0} \\
  & = & \frac{| \int \mathd^2 \tmmathbf{b}A ( s, \tmmathbf{b} ) ( b^2 \cos^2
  \phi ) |}{| \int \mathd^2 \tmmathbf{b}A ( s, \tmmathbf{b} ) |} 
  = \frac{\int \mathd b b^3 \Gamma ( s, b )}{2 \int \mathd b b \Gamma (
  s, b )} \quad,
\end{array}
\ee
where we have expanded the exponential and kept only the third
term. The first term does not depend on $q$ and the second term in the
expansion vanishes due to symmetry.

\begin{figure}
\begin{center}
\includegraphics[width=12cm]{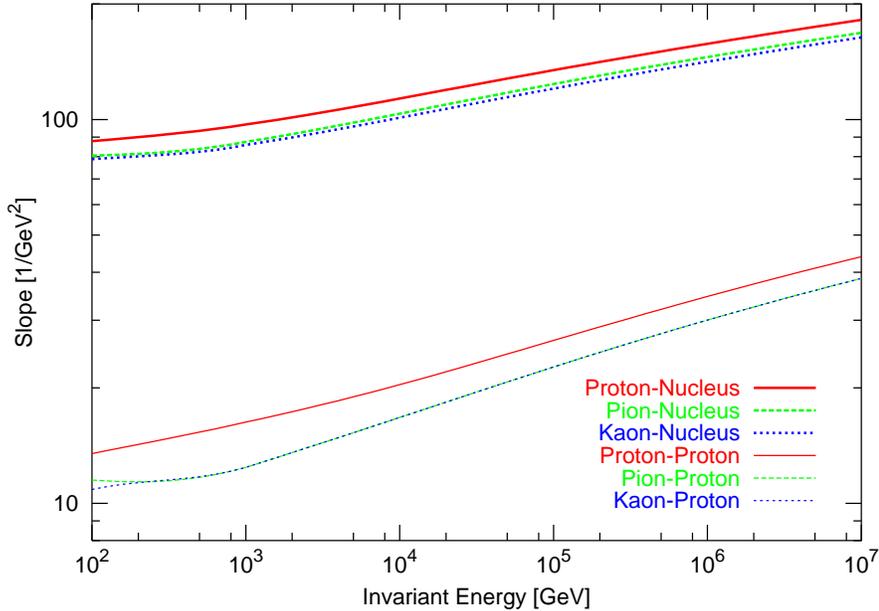}    % The printed column
\caption{The hadron-nucleus slopes (thin lines) and the hadron-nucleon slopes (thick lines) 
as a function of the collision energy in the center of mass frame.
Primary particles are Protons (full lines), $\pi$s (dashed lines), and Kaons (dotted lines).}
\label{pslope}                           
\end{center}  
\end{figure}

To obtain the scattering slope of a hadron-nucleus collision we replace the
hadron-nucleon scattering profile function by
\be
%\begin{array}{rcl}
  \Gamma_{\tmop{hA}} ( s, \tmmathbf{b} )  =  1 - \exp\left[- t \sum\limits^A_{i =
1}
  \chi_i ( s, \tmmathbf{b}_i )\right] 
%   =  1 - \prod^A_{i = 1} ( 1 - \Gamma_i ( s, \tmmathbf{b}_i ) )  
   =  1 - \left[ 1 - \tilde{T}_A ( \tmmathbf{b} ) \right]^A 
%\end{array}
\ee
where $\tilde{T}_A$ is obtained from the Glauber-Gribov
formalism \cite{Gribov:1968jf,Glauber:1970jm,Glauber:1987bb} by convolution of
the thickness function with the hadron-nucleon profile,
\be
\begin{array}{rcl}
  \tilde{T}_A ( \tmmathbf{b} ) & = & \int \mathd^2 \tmmathbf{c}\,T_A
  (\tmmathbf{b}) \Gamma (\tmmathbf{b}-\tmmathbf{c})\quad, \\
  T_A ( \tmmathbf{b} ) & = & \frac{1}{A} \int \mathd z\, \rho ( z,
\tmmathbf{b}
  ) . 
\end{array}
\label{eq_TAB} 
\ee
Using Eq. (\ref{eq_TAB}), the scattering slope for
elastic hadron-nucleus collisions Eq. (\ref{eq_slope}) is calculated as a function of
the collision energy. The underlying hadron-nucleon scattering slopes are taken 
from the SIBYLL model \cite{Fletcher:1994bd,Engel:1999db}.  
Fig. (\ref{pslope}) depicts the underlying hadron-nucleon slopes (thin lines) 
and the calculated  hadron-nucleus slopes (thick lines). The 
hadron-nucleus slopes are clearly higher than 
the hadron-nucleon slopes at the same energy. However, 
the ratios of the two slopes decreases with increasing energy. 
%%%%%%%%%%%%%%%%%%%%%%%%%%%%%%%%%%%%%%%%%%%%%%%%%%%%%%5

\section{Gravitational radiation from high energy cosmic rays}
\label{sec_CosmicRay}

After the derivation of the basic equations in the previous sections, we are now ready 
to calculate the amount of energy that is
emitted into gravitational radiation by a high energy proton propagating through the
atmosphere. 

The differential energy loss is given by
\begin{equation}\label{eqEn}
%\begin{displaymath}
\frac{dE}{dx}(s,d)=
\frac{\int\limits_0^{\sqrt{s}/2}dt\,
\frac{d \sigma^0_{\rm hA}}{d t}
E(t,s,d)}{\lambda \int\limits_0^{\sqrt{s}/2}dt \,
\frac{d\sigma^0_{\rm hA}}{d t}}\quad,
%\end{displaymath}
\end{equation}
where $\lambda$ is the mean free path for elastic scattering
 of the projectile in units of g/cm$^2$ and $\mathd \sigma^0_{\rm
 hA}/\mathd t$ is the differential 
hadron-nucleus cross section. 
For cosmic ray calculations it is convenient to calculate the energy loss $E$ 
in the laboratory frame. 
The corresponding Lorentz transformations are given in App. \ref{secLabSys}.

\begin{figure}
\begin{center}
\includegraphics[width=12cm]{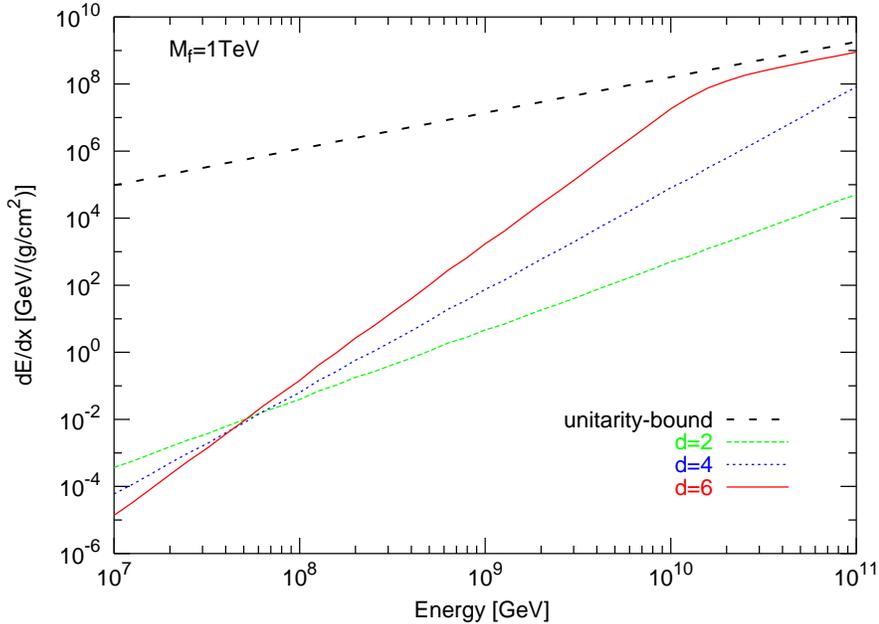}    % The printed column
\caption{Energy loss (in GeV/(g/cm$^2$)) of a proton propagating through
the atmosphere as a function of the lab-frame energy for $M_f=1$ TeV and $d=2,4,6$.}
\label{pdEdxMf1s2}                           
\end{center}  
\end{figure}

\begin{figure}
\begin{center}
\includegraphics[width=12cm]{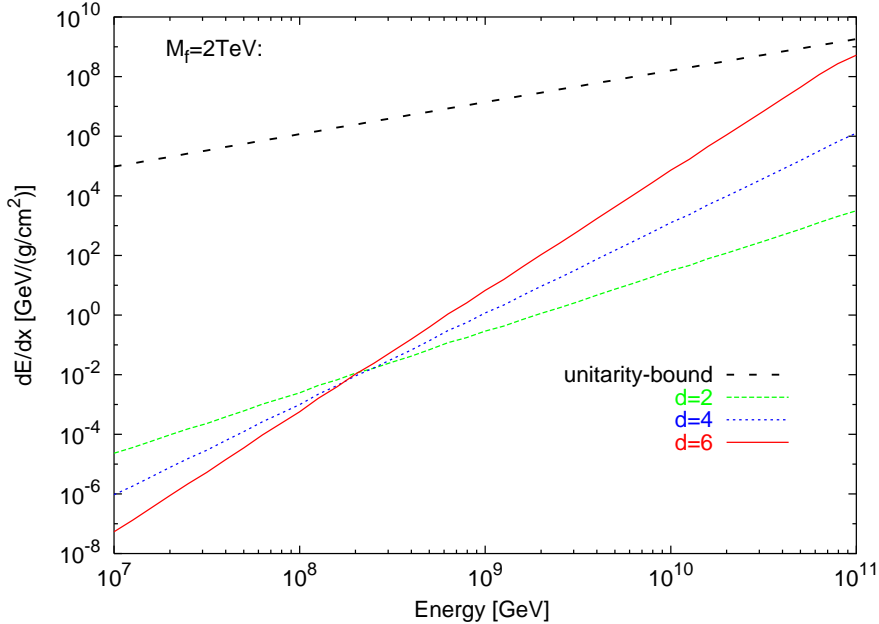}    % The printed column
\caption{Energy loss (in GeV/(g/cm$^2$)) of a proton propagating through
the atmosphere as a function of the lab-frame energy for $M_f=2$ TeV and $d=2,4,6$.}
\label{pdEdxMf2s2}                           
\end{center}  
\end{figure}

Figs. \ref{pdEdxMf1s2} (for $M_f=1$~TeV) and \ref{pdEdxMf2s2} (for $M_f=2$~TeV) show the differential energy loss
of a Proton propagating through the atmosphere as a function of the initial energy
in the laboratory frame. The short dashed, dotted and full lines give the results for two, four and six extra dimensions,
the long dashed lines show the unitarity bounds.
For large initial energies, a higher number of extra dimensions
leads to an enhancement  of the gravitational energy loss. 
However, with increasing fundamental scale $M_f$ the effect is much
weaker as shown in Fig. \ref{pdEdxMf2s2}.
Note, that the result is  cut-off
dependent as  $k_{\rm max}$ is not determined from first principles.
For the present study, we have chosen $k_{\rm max}=\sqrt{s}/2$,
which is the maximal value consistent with
energy conservation in the picture of
a gravitational wave being emitted by one of
the outgoing states.
The comparison of these results with \cite{Kazanas:2001ep} shows that
an approximation of the effects of extra dimension with a simple
phase space argument does yield a similar shape for the energy loss as those shown
in Figs. \ref{pdEdxMf1s2} and \ref{pdEdxMf2s2}. However, the omission of the correct kinematics
of the energy loss Eq. (\ref{eq_EvonTd246}) results in a dramatic 
overestimation of the gravitational energy loss effect by several orders of magnitude.  
In addition, the simple extension of the standard formula with  a modified phase space factor on the
integrated cross sections results in a violation of the unitarity
bound. 

Even though the energy loss into gravitational waves in our
(very optimistic) scenario is much lower than expected from previous
approximations, it might still have observable consequences for very high
energy  cosmic rays.  Therefore we implemented
Eq. (\ref{eq_EvonTd246}) and the elastic cross sections into a complete
cosmic ray air shower simulation (SENECA) \cite{Bossard:2000jh,Drescher:2002cr} to study
the modifications of the shower properties in detail.
  
\begin{figure}
\begin{center}
\includegraphics[width=12cm]{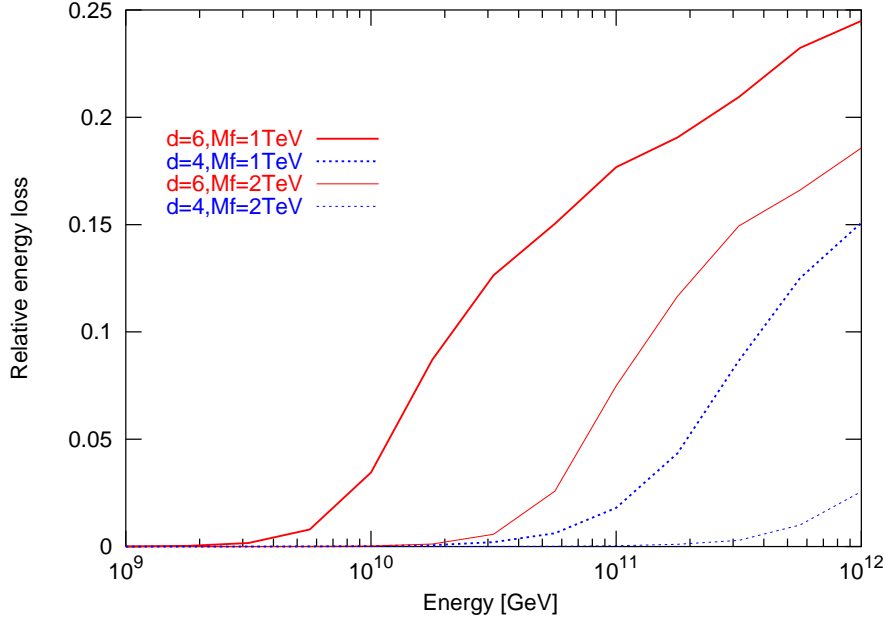}    % The printed column
\caption{Relative energy loss into gravitational radiation
as a function of the incident cosmic ray energy $E$
for $d=(4,6)$ and $M_f=(1,\,2)$~TeV.}
\label{pElossMf12s2}                           
\end{center}  
\end{figure}

Fig. \ref{pElossMf12s2} gives the relative energy loss as a function of the
incident energy $E$. The calculation is averaged over incident zenith 
angles $\rm d\!\cos(\theta)$ in the range $0^\circ \le \theta \le 60^\circ$.
The full lines indicate the calculations with six extra dimensions, while the dotted lines
show the results for four extra dimensions ($M_f=1$~TeV is shown by thick lines, $M_f=2$~TeV is shown by thin lines).  
For the case of two extra dimensions, deviations from the non-modified shower properties are very small even
for the most optimistic cases.
However, for  four extra dimensions
first deviations from the standard calculation become visible at energies higher than $5\cdot10^{10}$~GeV.
For $d=6$ the gravitational radiation becomes sizeable and already leads to deviations
around $5\cdot10^{9}$~GeV. At the highest energies, the integrated relative energy loss due to
gravitational radiation might even exceed 20\% of the initial particle energy.

In present day experiments, e.g. AUGER,  this gravitational energy loss
would show up as a decrease in the number of observed secondary particles.
The multiplicity of secondary particles $N_{\rm sec}(E,x)$
is directly observable in fluorescence experiments and is a key observable to
estimate the cosmic ray's initial energy. 
Any non-visible energy emission results in an underestimation of the initial energy 
in the energy reconstruction procedure. Thus, it has an impact on the interpretation of the
measured cosmic ray flux in dependence of the incoming particle energy.

How big is the distortion of the reconstructed flux due to graviton emission quantitatively?
Neglecting fluctuations, for a given incoming flux
$F={\rm{d}N}/{\rm{d}E}$,
the measured flux $F^\prime={\rm{d}N^\prime}/{\rm{d} E^\prime}$ 
depends on the
reconstructed energy $E^{\prime}(E)$. By identifying
the integrated fluxes $N^\prime(E^\prime)=N(E)$ one finds
\be
F^\prime(E^\prime) = \frac{\rm d N^\prime}{\rm d E^\prime}= 
\frac{\rm{d}N\left( E \right)}{\rm{d}E} \cdot 
\frac{\rm d E}{\rm d E^\prime} = F(E)\cdot \frac{\rm d E}{\rm d E^\prime} ~.
\ee
\begin{figure}
\begin{center}
\includegraphics[width=12cm]{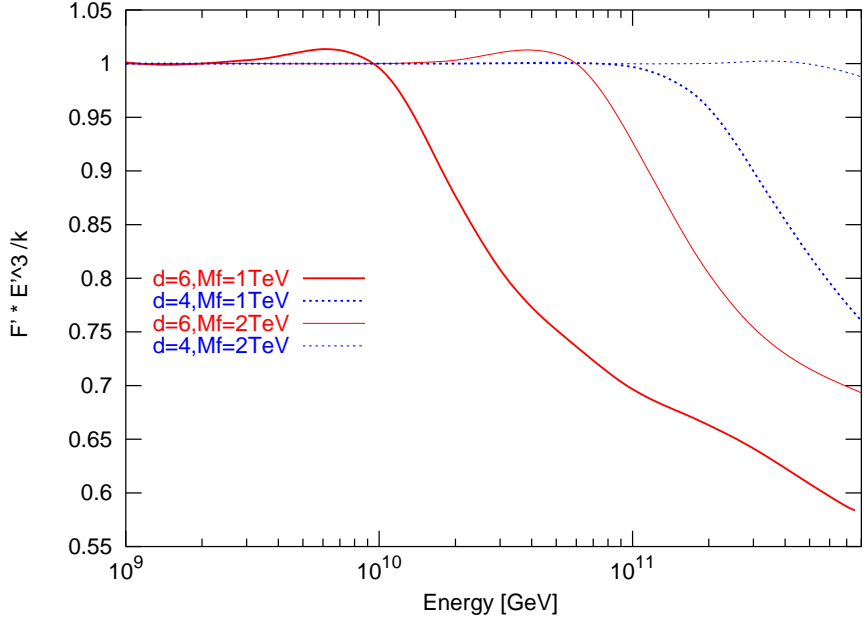} 
\caption{Reconstructed flux $F^{\prime} E^{\prime 3} /k $
as a function of the incident cosmic ray energy $E^{\prime}$
for $d=(4,6)$ and $M_f=(1,\,2)$~TeV.}                           
\label{pRecFlux}
\end{center}  
\end{figure}

For an incoming flux  $F=k E^{-3}$ the flux reconstruction is
shown in Fig. \ref{pRecFlux}.  In all scenarios ($d\ge
4$) gravitational wave emission might indeed influence the energy
reconstruction above $5\cdot10^9$~GeV.
For ultra high energy cosmic rays even an apparent cut-off seems possible\footnote{Note the linear 
scale on the y-axis, thus the suppression does not mimic the GZK cut-off.}%
, because the relative amount of non-visible energy increases strongly with increasing energy. 
Hence, for UHECRs the interpretation of experimental
data might have to be modified in scenarios with large
extra dimensions.

Presently available data from Hires and AGASA do not allow one to
observe the predicted suppression pattern, because even in our most
optimistic scenario the flux is reduced only by a factor of 0.5 for
the highest energies. However, with the expected high statistics data
from the Pierre Auger Observatory a detailed exploration of this
phenomenon might be possible.

As a remark, we want to point out that in our calculation, gravitational wave emission does not 
give new insights into phenomena at lower energies ($E\le 10^{18}$~eV) and can not be
considered as a candidate to explain the famous knee in the cosmic ray spectrum.

%%%%%%%%%%%%%%%%%%%%%%%%%%%%%%%%%%%%55
\section{Conclusion}

The energy loss into gravitational waves is calculated for ultra high
energy cosmic rays.  In contrast to previous estimates, quasi-elastic
particle scattering in the ADD scenario with 4 or 6 extra dimensions
has no observable influence on the properties of cosmic ray air
showers at incident energies below $5\cdot10^{18}$~eV.  Thus, the
emission of gravitational radiation can not be used to explain the
steepening of the cosmic ray spectrum at the ''knee'' ($E\sim
10^{15.5}$~eV).  For two large extra dimensions, the studied effects
are generally too small to lead to any observable effect.

However, for energies above $5\cdot10^{18}$~eV  and $M_f\le 2$~TeV, $d\ge 4$
gravitational energy loss during the air shower evolution can be sizeable.
This might result in an underestimation of the reconstructed energy for 
ultra high energy cosmic rays as studied by Hires, AGASA and
 the Pierre Auger Observatory.

%%%%%%%%%%%%%%%%%%%%%%%%%%%%%%%%%%%%%%%%%%%%%%%%%
\begin{ack}             
The authors thank Drs.
S. Hossenfelder, S. Hofmann and S. Ostaptchenko for fruitful discussions. B.K. thanks
the Frankfurt International Graduate School of Science (FIGSS)
for financial support through a PhD fellowship. This work was supported by GSI and BMBF.

All numerical calculations have been performed at the
Frankfurt Center for Scientific Computing (CSC).
\end{ack}

%\bibliographystyle{plain}        % Include this if you use bibtex 
%\bibliography{autosam}           % and a bib file to produce the 
                                 % bibliography (preferred). The
                                 % correct style is generated by
                                 % Elsevier at the time of printing.

%%%%%%%%%%%%%%%%%%%%%%%%%%%%%%%%%%%%%%%%%%%%%%%%5
\begin{appendix}
\section{Energy loss in the lab system}
\label{secLabSys}

Equation (\ref{eq_EvonTd246}) provides the 
gravitationally radiated energy in the centre of mass frame of the reaction. 
To transform the kinematic variables
to the laboratory frame with a target Proton at rest one has
to apply the Lorentz transformation matrix
\begin{equation}
\label{eq_LorentzTrafo1}
\Lambda=
\left(
\begin{array}{cc}
\rm{Cosh} (\eta) & \rm{Sinh} (\eta)\\
\rm{Sinh} (\eta) & \rm{Cosh} (\eta)
\end{array}
\right)=
\left(
\begin{array}{cc}
\frac{\sqrt{s}}{2 m_p} & \frac{\sqrt{s-16 m_p^2}}{2 m_p}\\
\frac{\sqrt{s-16 m_p^2}}{2 m_p} & \frac{\sqrt{s}}{2 m_p}
\end{array}
\right)\quad,
\end{equation}
which acts on the 
$t$ and the $z$ (i.e. longitudinal) component of the $4+d$ dimensional vector. All
the other (transverse) components remain unchanged.
Eq. (\ref{eq_EvonTd246}) gives
the energy $E$ and momentum $\ul{k}$ of the gravitational radiation emitted from one of
the interacting particles ($p_1,\,p_2$). 
For different momentum directions 
${\ul{k}}/{\left|\ul{k}\right|}$
the Lorentz transformation Eq. (\ref{eq_LorentzTrafo1})
gives  different energy losses in the lab-frame. 
To avoid this complication we use  a mean value of
the left over four momentum $\overline{p^{\prime}}$
of the scattering particles.
If the energy is radiated away from particle
$i$ we define $p_i^{\prime}=p_i-k$.
Averaging over these cases yields
\begin{equation}
\label{eq_Pbar1}
\overline{p^{\prime}}_{\rm}=
\sum_i^N \frac{p_i^{\prime}+\sum_{l\neq i}^N p_l}{N}.
\end{equation}
Using the symmetry of a $2\rightarrow 2$ scattering
in the centre of mass system we find
\begin{equation}
\label{eq_Pbar2}
\overline{p^{\prime}}_{\rm{CM}}=
\left(
\sqrt{s}-k_0,0,0,0,\dots
\right).
\end{equation}
Because $\overline{p^{\prime}}_{\rm{CM}}$ has
no $z$ component, the mean left over
energy in the laboratory system becomes
\begin{equation}
\label{eq_PbarLab}
\overline{p^{\prime}}_{\rm{lab}}=
\Lambda \cdot\overline{p^{\prime}}_{\rm{CM}}
=\left(
\frac{\sqrt{s}}{2 m_p}(\sqrt{s}-k_0)
,0,0...
\right).
\end{equation}
From Eq. (\ref{eq_PbarLab})
the mean energy loss in the lab system is obtained as
\begin{equation}
\overline{E}_{\rm lab}^{\rm loss} =\frac{s}{2 m_p}-\overline{p^{\prime}}_{0\,\rm{lab}}
=\frac{\sqrt{s}}{2 m_p}
E^{\rm loss}_{\rm CM}.
\end{equation}

\end{appendix}

\appendix
\end{document}